\documentclass[11pt]{article}
\usepackage{moriond}
\usepackage{amsmath}
\usepackage{amsfonts}
\usepackage{amssymb}
\usepackage{graphicx}

\newcommand{\Photo}{\includegraphics[height=35mm]{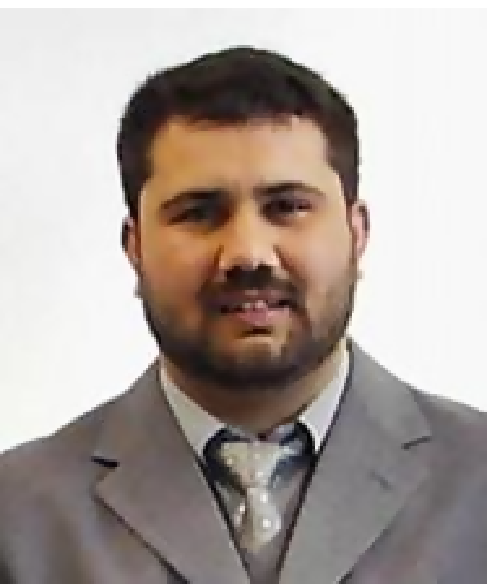}}

\begin{document}
\title{Three-Loop Neutrino Mass Models at Colliders}
\author{Amine Ahriche $^{1,2}$, Kristian~L. McDonald $^{3}$ \& Salah Nasri $^{4}$}

\address{$^{1}$ Department of Physics, University of Jijel, PB 98
Ouled Aissa, DZ-18000 Jijel, Algeria\\
$^{2}$ The Abdus Salam International Centre for Theoretical Physics,
Strada Costiera 11,\\
 I-34014, Trieste, Italy\\
$^{3}$ ARC Centre of Excellence for Particle Physics at the
Terascale, School of Physics,\\
The University of Sydney, NSW 2006, Australia.\\
$^{4}$ Physics Department, UAE University, POB 17551, Al Ain, United
Arab Emirates}

\maketitle\abstracts{ In this work, we report on recent analyses of
a class of models that generate neutrino mass at the three-loop
level. We argue that these models offer a viable solution to both
the neutrino mass and dark matter problems, without being in
conflict with experimental constraints from, e.g. lepton flavor
violating processes and the muon anomalous magnetic moment.
Furthermore, we describe observable experimental signals predicted
by the models and show that they have common signatures that can be
probed at both the LHC and ILC.}

\section{Introduction}

The Standard Model has been remarkably successful in describing physics at the
weak scale. However, many questions remain, including those relating to the
origin of neutrino mass and the reason for its smallness. In this context,
models with radiative neutrino mass are of significant experimental interest.
These models provide an inherent loop-suppression that allows the new physics
responsible for neutrino mass to be lighter than in other scenarios. This loop
suppression is more severe as the number of loops increases, making models
with three-loop masses particularly interesting, as they generically require
new physics near the TeV scale. Such light new particles can be produced and
detected within current and near-future experiments by searching for
signatures such as lepton flavor violating (LFV) effects.

Here we present a class of models with common features, in which neutrino mass
is generated at the three-loop
level~\cite{KNT,Ahriche:2014cda,Ahriche:2014oda,Ahriche:2015wha,Chen:2014ska},
and discuss interesting signatures that can appear at both leptonic and
hadronic colliders. We focus primarily on the KNT model~\cite{KNT} and present
recent analyses showing that the model satisfies LFV constraints, such as
$\mu\rightarrow e+\gamma$, and fits the neutrino oscillation data.
Furthermore, the model contains a viable candidate for the dark matter (DM) in
the universe, in the form of a light right-handed (RH) neutrino. We also show
that a strongly first order electroweak phase transition can be achieved with
a Higgs mass of $\simeq125$ GeV, as measured at the LHC~\cite{ATLAS,CMS}. The
model contains new charged scalars that may lead to significant modification
on the Higgs decay channel $h\rightarrow\gamma\gamma$\ while $h\rightarrow
\gamma\gamma$\ remains SM-like. We also discuss possible signature of this
class of models at the ILC and LHC through possible modifications of the
processes $e^{-}e^{+}\rightarrow e^{-}\mu^{+}+E_{miss}$ \ and $pp\rightarrow
e^{-}e^{+}+E_{miss}$, $\mu^{-}\mu^{+}+E_{miss}$, $e^{-}\mu^{+}+E_{miss}$ respectively.

\section{A Class of Three-Loop Models}

The class of models we discuss is based on the KNT model~\cite{KNT}, which is
obtained by extending the SM to include three right-handed (RH) Majorana
neutrinos and two electrically charged scalars, $S_{1}^{\pm}$ and $S_{2}^{\pm
}$, all of which are singlets under $SU(2)_{L}$. In addition, a discrete
$Z_{2} $ symmetry is imposed, under which $\{S_{2},N_{i}\}\rightarrow
\{-S_{2},-N_{i}\}$, and all other fields are even. The generalized class of
models is obtained by promoting the charged scalar $S_{2}^{\pm}$ to a scalar
multiplet $T$, and the three RH neutrinos $N_{i}$ to three generations of
fermionic multiplets $E_{i}$, while retaining the same charges under the
$Z_{2}$ symmetry \footnote{Except for the septuplet case where the global
symmetry $Z_{2}$ is accidental \cite{Ahriche:2015wha}.}. This symmetry plays
two key roles, preventing a tree-level coupling between $N_{R}$ ($E_{i}$) and
the SM Higgs, which would otherwise induce tree-level neutrino masses, and
ensuring that the lightest neutral fermion $E_{i}^{0}$ is a stable DM
candidate. The general Lagrangian reads as
\begin{equation}
\mathcal{L}=\mathcal{L}_{SM}+\{f_{\alpha\beta}L_{\alpha}^{T}Ci\tau_{2}%
L_{\beta}S_{1}^{+}+g_{i\alpha}\bar{E}_{i}T\ell_{\alpha R}-\frac{1}{2} \bar
{E}_{i}^{c}M_{ij}E_{j}+h.c\}-V,\label{L}%
\end{equation}
where $L_{\alpha}$ is the left-handed lepton doublet, $f_{\alpha\beta}$ are
Yukawa couplings which are antisymmetric in the generation indices $\alpha$
and $\beta$, $M_{ij}$\ are the fermionic mass matrix elements, $C$ is the
charge conjugation matrix, and $V(\Phi,S_{1},T)$ is the tree-level scalar
potential. Here $\Phi$ denotes the SM Higgs doublet.

\begin{figure}[t]
\begin{centering}
\includegraphics[width=7cm,height=2.5cm]{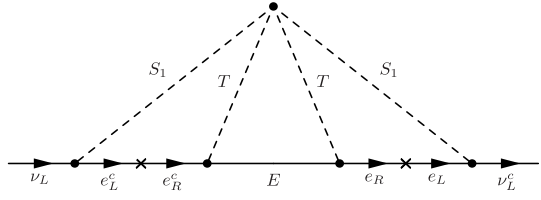}
\par\end{centering}
\caption{\textit{The three-loop diagram that generates the neutrino mass.}}%
\label{diag}%
\end{figure}

Using interactions in (\ref{L}) together with the scalar interaction
$V\supset\lambda_{s}S_{1}^{+}S_{1}^{-}T^{\dag}T$, the neutrino mass matrix
elements can arise from the three-loop diagram in Fig. \ref{diag}, that are
given by \cite{AN2013}
\begin{equation}
(M_{\nu})_{\alpha\beta}=\frac{(2n+1)\lambda_{s}m_{\ell_{i}}m_{\ell_{k}}%
}{\left(  4\pi^{2}\right)  ^{3}M_{T}}f_{\alpha i}f_{\beta k}g_{ij}%
g_{kj}F\left(  M_{E_{j}}^{2}/M_{T}^{2},M_{S_{1}}^{2}/M_{T}^{2}\right)
,\label{nu-mass-1}%
\end{equation}
where $\rho,\kappa(=e,\mu,\tau)$ are the charged leptons flavor indices,
$i=1,2,3$ denotes the three $E_{i}$ multiplets, and the function $F$ is a loop
integral which is $\mathcal{O}(1)$~\cite{AN2013}. It is interesting to note
that, unlike the conventional seesaw mechanism, the radiatively generated
neutrino masses are directly proportional to the charged lepton and RH
neutrino masses, as well as being loop-suppressed. Here $n=0$ corresponds to
the KNT model, while $n=1,2,3$ gives generalizations where $E_{i}$ and $T$ are
$SU(2)_{L}$ triplets, quintuplets and septuplets, respectively (i.e.~$T$ and
$E_{i}$ are both assigned to the $(2n+1)$ representation under $SU(2)_{L}$ and
carry two units of hypercharge).

The Lagrangian (\ref{L}) induces flavor violating processes, such as
$\ell_{\alpha}\rightarrow\gamma\ell_{\beta}$ for $m_{\ell_{\alpha}}%
>m_{\ell_{\beta}}$, and an extra contribution to the muon anomalous magnetic
moment. Both are generated at one loop via the exchange of the
charged scalar $S_{1}^{\pm}$, and the members of the multiplets $T$
and $E_{i}$. The LFV branching ratios and the muon anomalous
magnetic moment are given by

\begin{align}
B(\ell_{\alpha}  \rightarrow\gamma\ell_{\beta})=\frac{3\alpha_{em}%
\upsilon^{4}}{36\pi}\left\{  \frac{\left\vert
f_{\kappa\alpha}f_{\kappa\beta
}^{\ast}\right\vert ^{2}}{36M_{S_{1}}^{4}}+\frac{\left(  2n+1\right)  ^{2}%
}{M_{T}^{4}}\left\vert {\sum\limits_{i}}g_{i\alpha}g_{i\beta}^{\ast}%
F_{2}\left(  M_{E_{i}}^{2}/M_{T}^{2}\right)  \right\vert
^{2}\right\},\label{mutoegamma}\\
\delta a_{\mu} =-\frac{m_{\mu}^{2}}{16\pi^{2}}\left\{
\sum_{\alpha\neq
\mu}\frac{|f_{\mu\alpha}|^{2}}{6M_{S_{1}}^{2}}+n\sum_{i}\frac{|g_{i\mu}|^{2}%
}{M_{T}^{2}}F_{2}(M_{E_{i}}^{2}/M_{T}^{2})\right\},
\end{align}
with $\kappa\neq\alpha,\beta$, $\alpha_{em}$ being the fine
structure constant and $F_{2}(x)=(1-6x+3x^{2}+2x^{3}-6x^{2}\ln
x)/6(1-x)^{4}$.

In our scan of the parameter space of the model we impose the experimental
bounds on $B(\mu\rightarrow e\gamma)$~\cite{LFV}, $B(\tau\rightarrow\mu
\gamma)$\ and $\delta a_{\mu}$~\cite{PDG}, and use the allowed values for the
neutrino mixing parameters, $s_{12}^{2}=0.320_{-0.017}^{+0.016}$, $s_{23}%
^{2}=0.43_{-0.03}^{+0.03}$, $s_{13}^{2}=0.025_{-0.003}^{+0.003}$, and the mass
squared differences, $\left\vert \Delta m_{31}^{2}\right\vert =2.55_{-0.09}%
^{+0.06}\times10^{-3}$ \textrm{eV}$^{2}$ and $\Delta m_{21}^{2}=7.62_{-0.19}%
^{+0.19}\times10^{-5}\mathrm{eV}^{2}$ ~\cite{GF}.

\section{Dark Matter}

An immediate implication of the $Z_{2}$ symmetry is that that lightest neutral
fermion, $E_{1}^{0}$, is stable, and hence a candidate for dark matter (DM).
The $E_{1}^{0}$ number density gets depleted through the process $E_{1}%
^{0}E_{1}^{0}\rightarrow\ell_{\alpha}\ell_{\beta}$ via the $t$- and
$u$-channel exchange of $T$. In the singlet case ($n=0$), the non-relativistic
limit of the annihilation cross section gives
\begin{equation}
\sigma_{E_{1}^{0}E_{1}^{0}}\upsilon_{r}\simeq\sum_{\alpha,\beta}|g_{1\alpha
}g_{1\beta}^{\ast}|^{2}\frac{M_{E_{1}}^{2}\left(  M_{T}^{4}+M_{E_{1}}%
^{4}\right)  }{48\pi\left(  M_{T}^{2}+M_{E_{1}}^{2}\right)  ^{4}}\upsilon
_{r}^{2},\label{sig11}%
\end{equation}
with $\upsilon_{r}$ is the relative velocity between the annihilation
$E_{1}^{0}$'s. In cases with nontrivial representations ($n\neq0$), there
exist other annihilation channels, such as $E_{1}^{0}E_{1}^{0}\rightarrow WW$,
which increase the animation cross section, and therefore the DM candidate
should be heavier. The WW annihilation cross section contribution is given by%
\begin{equation}
\sigma_{E_{1}^{0}E_{1}^{0}\rightarrow WW}\upsilon_{r}=\frac{\pi\alpha_{2}^{2}%
}{M_{E_{1}}^{2}}\left(  a+b\upsilon_{r}^{2}\right)  ,
\end{equation}
with the $SU(2)_{L}$ structure constant $\alpha_{2}=g^{2}/4\pi$; and \{$a,b
$\}=\{$\frac{37}{12},\frac{17}{48}$\}, \{$\frac{207}{20},\frac{243}{80}$\},
\{$\frac{174}{7},\frac{263}{28}$\} for $n=1,2,3$ respectively.

When combining the relic density together with the neutrino mass and mixing,
LFV and muon anomalous magnetic moment bounds, the mass of the charged scalar
$S_{1}$ should exceed 100 GeV, while the bounds on $E_{i}$ and $T$ are
sensitive to the $SU(2)_{L}$ quantum numbers. For the KNT case (n=0), we find
that $M_{E_{1}^{0}}<225$ \textrm{GeV}\ while $M_{T}<245$ \textrm{GeV}%
~\cite{AN2013}. For the triplet, quintuplet and septuplet cases the DM mass
should be in the ranges $M_{E_{1}^{0}}=2.35\sim2.75$
TeV~\cite{Ahriche:2014cda}, $M_{E_{1}^{0}}\sim6$ TeV~\cite{Ahriche:2014oda}
and $M_{E_{1}^{0}}\sim21$ TeV~\cite{Ahriche:2015wha} respectively, with
$M_{T}>M_{E_{1}^{0}}$.

\section{Electroweak Phase Transition}

Although the SM has all the qualitative ingredients for electroweak
baryogenesis, the amount of matter-antimatter asymmetry generated is too
small. One of the reasons for this smallness is the fact that the electroweak
phase transition (EWPT) is not strongly first order, which is necessary to
suppress the sphaleron processes in the broken phase. The EWPT strength can be
improved if new scalar degrees of freedom around the electroweak scale are
coupled to the SM Higgs, which is the case in this class of models.

The investigation of the scalar effective potential reveals that, within the
allowed parameter space of the model, the strength of the electroweak phase
transition (EWPT) can be first order~\cite{AN2013}. We found that if the
one-loop corrections to the Higgs mass are sizeable, then the strongly first
order EWPT condition, $\upsilon(T_{c})/T_{c}>1$, can be realized while keeping
the Higgs mass around $125$~GeV. The reason for this being that the extra
charged singlets affect the dynamics of the SM scalar field VEV around the
critical temperature~\cite{hna}.

\begin{figure}[h]
\includegraphics[width=7.5cm,height=5.5cm]{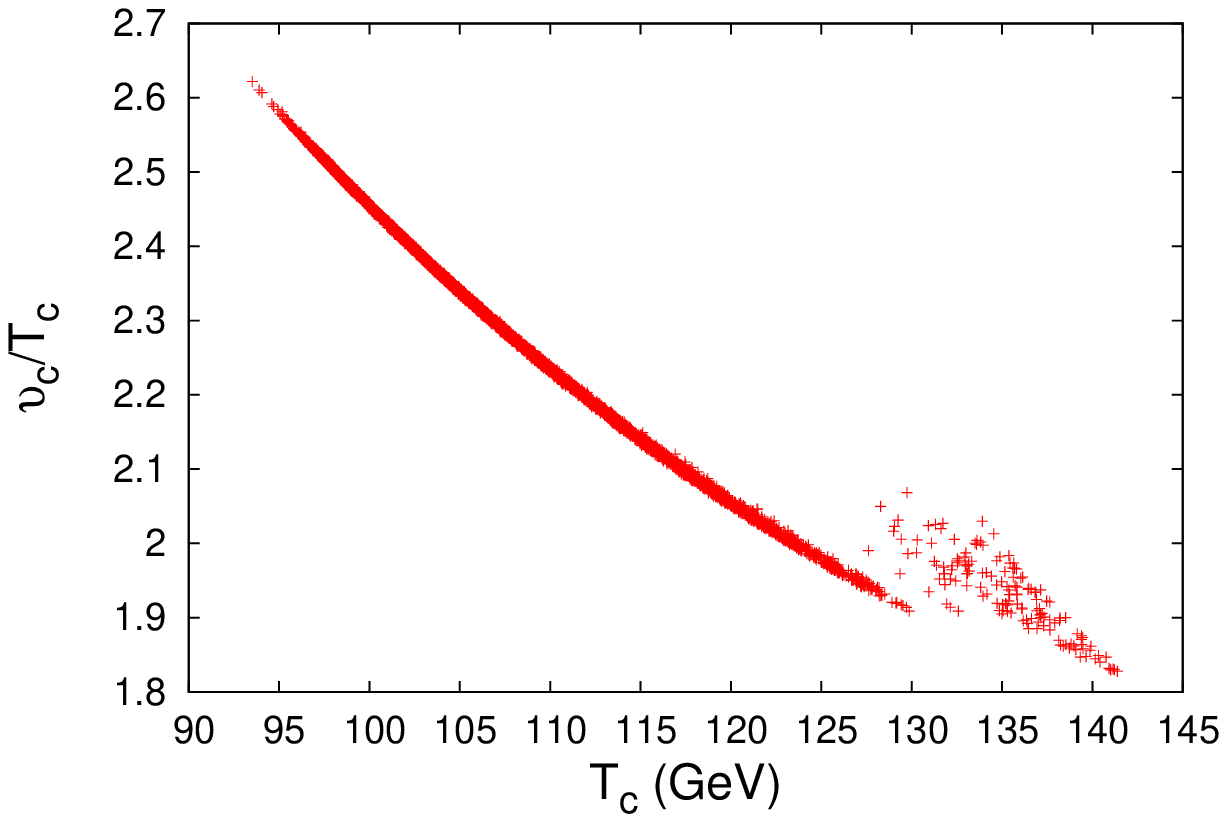}~\includegraphics[width=8cm,height=5.5cm]{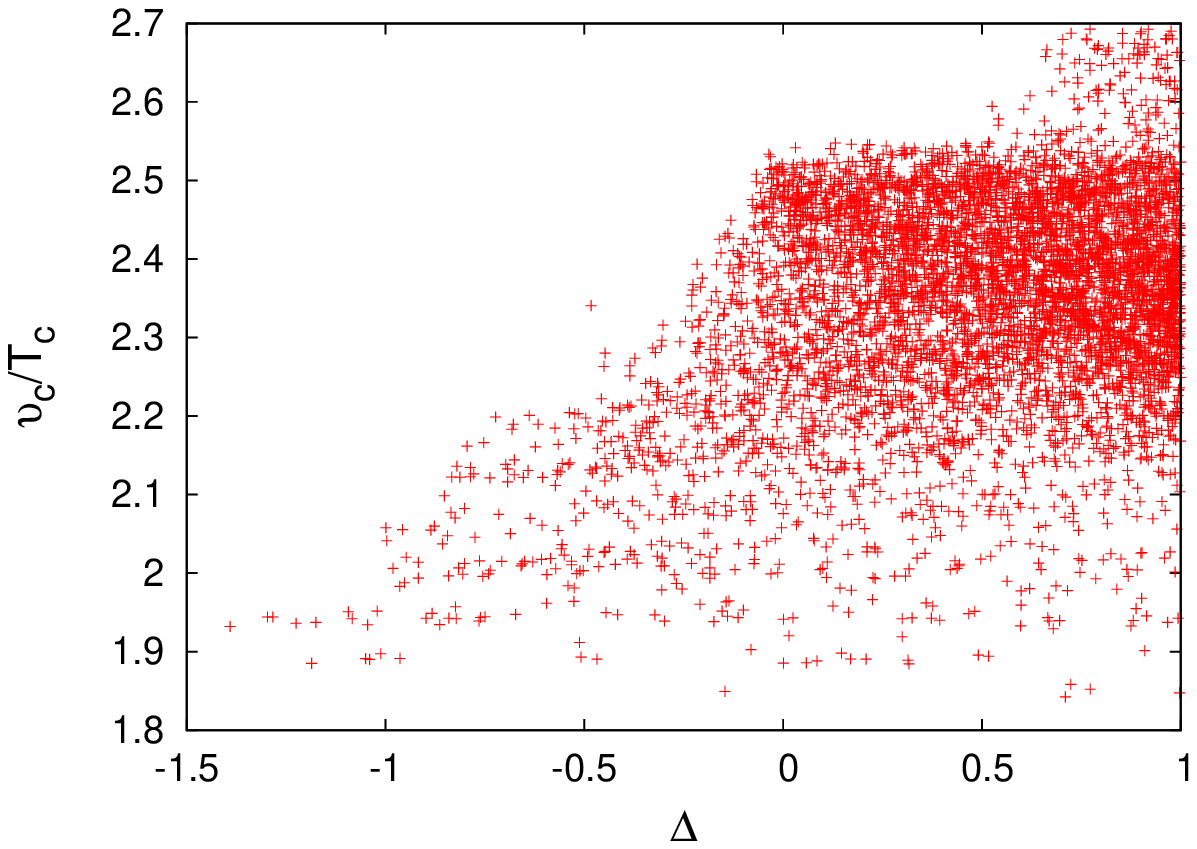}\caption{\textit{Left:
the critical temperature is presented versus the quantity $\upsilon_{c}/T_{c}%
$. Right: the ratio $\upsilon_{c}/T_{c}$ versus the relative enhancement in
the Higgs triple coupling $\Delta=\left(  \lambda_{hhh}^{(3)}-\lambda
_{hhh}^{SM}\right)  /\lambda_{hhh}^{SM}$.}}%
\label{ct}%
\end{figure}

The existence of extra fields coupled to SM Higgs doublet will induce one-loop
corrections to the triple Higgs coupling, $\lambda_{hhh}^{(3)}$, which is of
great interest, especially at leptonic colliders. In Fig.~\ref{ct}-left, we
show the plot for $\upsilon(T_{c})/T_{c}$ versus the critical temperature. One
observes that a strongly first order EWPT is possible while the critical
temperature lies around 100 $\mathrm{GeV}$. In Fig.~\ref{ct}-right, we show
the ratio $\upsilon(T_{c})/T_{c}$ versus the relative enhancement on the
triple Higgs coupling due to new physics, $\Delta=\left(  \lambda_{hhh}%
^{(3)}-\lambda_{hhh}^{SM}\right)  /\lambda_{hhh}^{SM}$. It is clear that the
enhancement is significant when the EWPT is stronger.

\section{Collider Phenomenology}

In these models, there are many common signatures at both the ILC and LHC.
Here, we briefly discuss two common signals, one at the ILC and another at the
LHC. At the ILC, the process $e^{-}e^{+}\rightarrow e^{-}\mu^{+}+E_{miss}$ is
modified in this class of models, where $E_{miss}\equiv\nu_{\mu}\bar{\nu}_{e}%
$, $\nu_{e}\bar{\nu}_{\tau}$, $\nu_{\tau}\bar{\nu}_{e}$,
$\nu_{\mu}\bar{\nu }_{\mu}$, $\nu_{\tau}\bar{\nu}_{\mu}$,
$\nu_{\tau}\bar{\nu}_{\tau}$, $E^{0}E^{0}$. The first six
combinations are mediated by $W^{\pm}$ or/and $S_{1}^{\pm}$ while
those of $E^{0}E^{0}$ are mediated by $T^{\pm}$. Here $E^{0}$ could
be $E_{1}^{0}$ and possibly $E_{2,3}^{0}$ if it decays into a
charged lepton and $T^{\pm}$ outside the detector. The background is
given by the process $E_{miss}\equiv\nu_{\mu}\bar{\nu}_{e}$, which
occurs in the SM via 18 Feynamnn diagrams and via 40 diagrams in the
present class of models \cite{ANS}. The total expected cross section
and the expected number of events for the processes
$e^{-}e^{+}\rightarrow e^{-}\mu^{+}+E_{miss}$ are represented by
$\sigma^{EX}$ and $N^{EX}=\mathcal{L}\sigma^{EX}$, with
$\mathcal{L}$\ being the integrated luminosity. In the SM case we
have $N^{B}=\mathcal{L}\sigma^{B}$. As an example, we consider the
following benchmark for the KNT case (n=0)~\cite{ANS}
\begin{align}
f_{e\mu}  & =-(4.97+i1.41)\times10^{-2},~f_{e\tau}=0.106+i0.0859,~f_{\mu\tau
}=(3.04-i4.72)\times10^{-6},\nonumber\\
g_{i\alpha}  & =10^{-2}\times\left(
\begin{array}
[c]{ccc}%
0.2249+i0.3252 & 0.0053+i0.7789 & 0.4709+i1.47\\
1.099+i1.511 & -1.365-i1.003 & 0.6532-i0.1845\\
122.1+i178.4 & -0.6398-i0.6656 & -10.56+i68.56
\end{array}
\right)  ,\nonumber\\
M_{E_{i}^{0}}  & =\{162.2\mathrm{~GeV},\ 182.1\mathrm{~GeV}%
,~209.8\mathrm{~GeV}\},~M_{S_{1}}=914.2\mathrm{~GeV},~M_{T}=239.7\mathrm{~GeV}
\end{align}
We used CalcHEP~\cite{calc} to simulate the model and generate the
differential cross section and the relevant kinematic variables for
different CM energy: $E_{CM}=$250 , 350, 500 \textrm{GeV} and 1
\textrm{TeV}, initially with unpolarized beams; and then we consider
polarized beams with $P\left( e^{-},e^{+}\right)  =[-0.8,+0.3]$
and/or $P\left(  e^{-},e^{+}\right) =[+0.8,-0.3]$. After imposing
the appropriate cuts in both cases of polarized and unpolarized
beams, we summarize the results for the corresponding luminosity
values in Table-\ref{T1}. \begin{table}[h]
\begin{centering}
\begin{tabular}{|c|c|c|c|c|c|}
\hline\hline $E_{CM}$ $($\textrm{GeV}$)$ & $L$ $(fb^{-1})$ &
$P(e^{-},e^{+})$ & $N_{B}$ & $N_{EX}$ & $N_{S}$ \\ \hline $250$ &
$250$ & $0,0$ & $16480$ & $16851$ & $371$ \\ \cline{3-6} & &
$-0.8,+03$ & $38498$ & $39775$ & $1277$ \\ \hline $350$ & $350$ &
$0,0$ & $20609$ & $21055$ & $446$ \\ \cline{3-6} & & $-0.8,+03$ &
$47740$ & $48990$ & $1250$ \\ \hline $500$ & $500$ & $0,0$ & $28280$
& $28815$ & $535$ \\ \cline{3-6} & & $-0.8,+03$ & $65500$ & $67250$
& $1750$ \\ \hline $1000$ & $1000$ & $0,0$ & $19.217$ & $469.76$ &
$450.54$ \\ \cline{3-6} & & $+0.8,-03$ & $2.07$ & $727.10$ &
$725.03$ \\ \hline\hline
\end{tabular}
\par\end{centering}
\caption{The expected ($N_{EX}$) and background ($N_{B}$) number of events for
different CM energy values with/without polarized beams within the cuts given
in Table-\ref{T1}.}%
\label{T1}%
\end{table}

\begin{figure}[h]
\begin{centering}
\includegraphics[width=7.5cm,height=5.5cm]{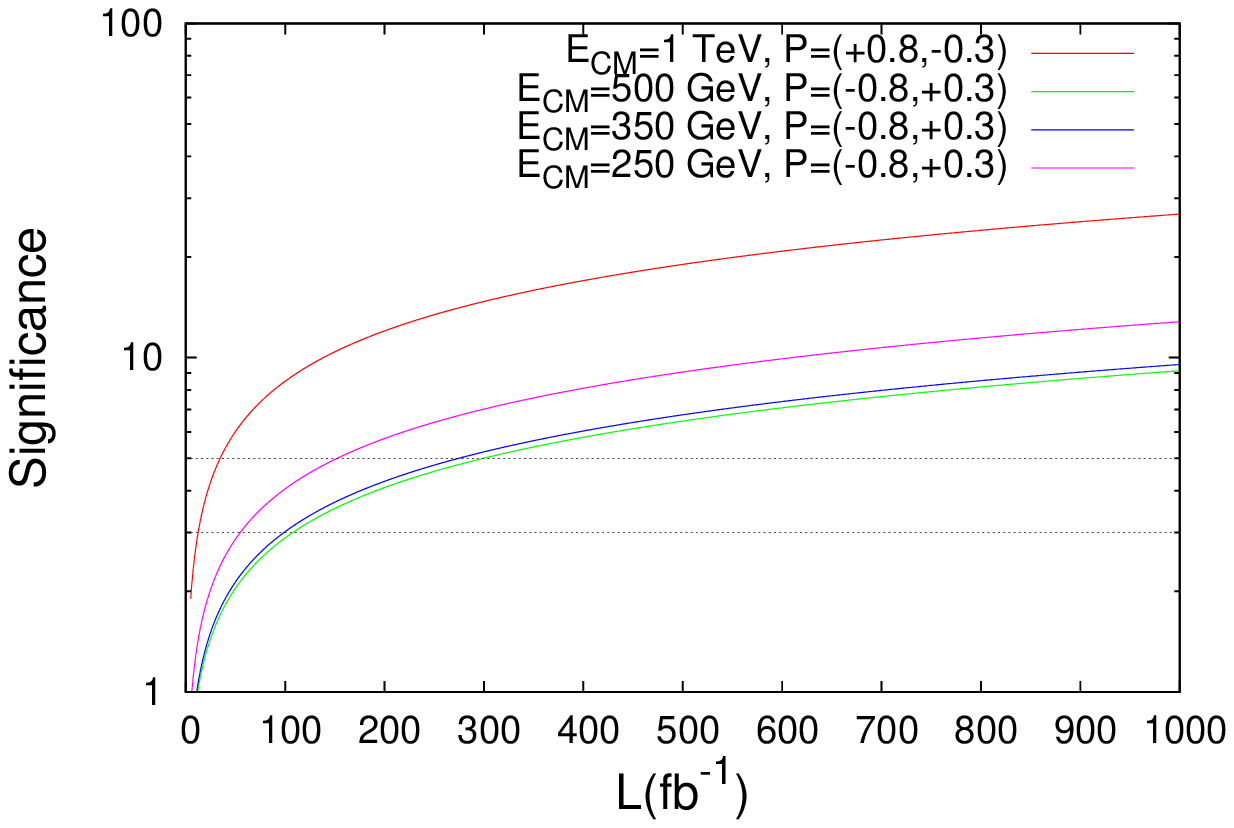}~\includegraphics[width=7.5cm,height=5.5cm]{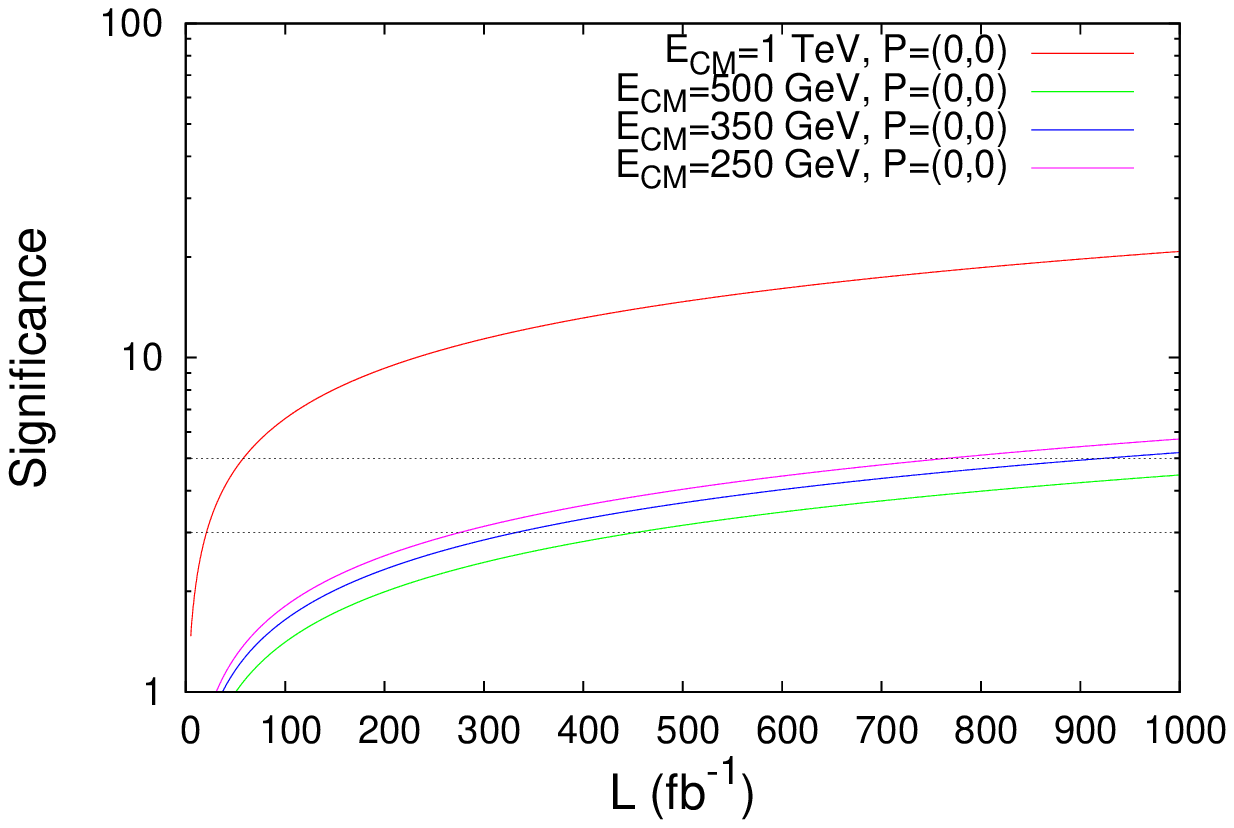}
\par\end{centering}
\caption{\textit{The significance versus luminosity at different CM energies
within the cuts defined in Table-\ref{T1}; with (left) and without (right)
polarized beams. The two horizontal dashed lines represent $\mathcal{S}=3$ and
$\mathcal{S}=5$, respectively.}}%
\label{SS}
\end{figure}

In Fig.~\ref{SS}, we show the dependance of the significance on the
accumulated luminosity with and without polarized beams for the
considered CM energies.
We clearly see that for a polarized beam, the signal can be
observed even with relatively low integrated luminosity. For example, at
$E_{CM}=250$ \textrm{GeV}, the 5$\sigma$ required luminosity is 150 $fb^{-1}$
for polarized beam as compared to 700 $fb^{-1}$ without polarization.

\begin{figure}[h]
\begin{centering}
\includegraphics[width=7.5cm,height=5.5cm]{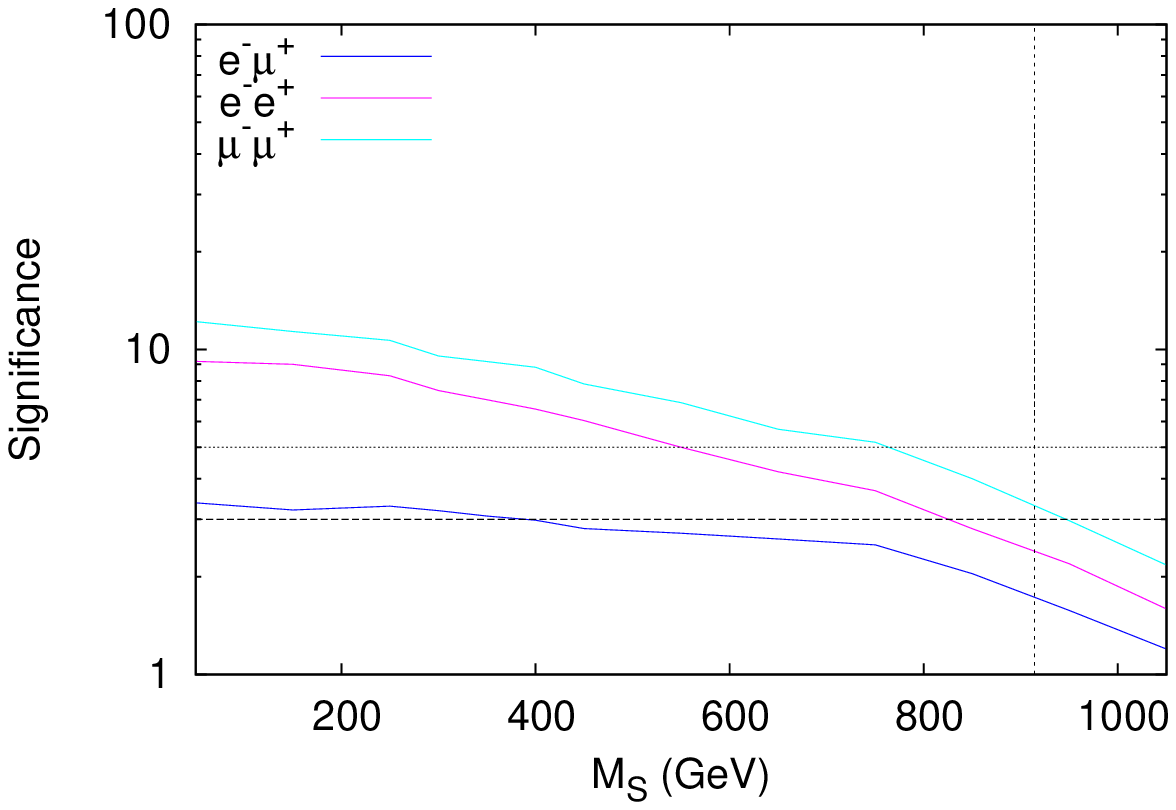}~\includegraphics[width=7.5cm,height=5.5cm]{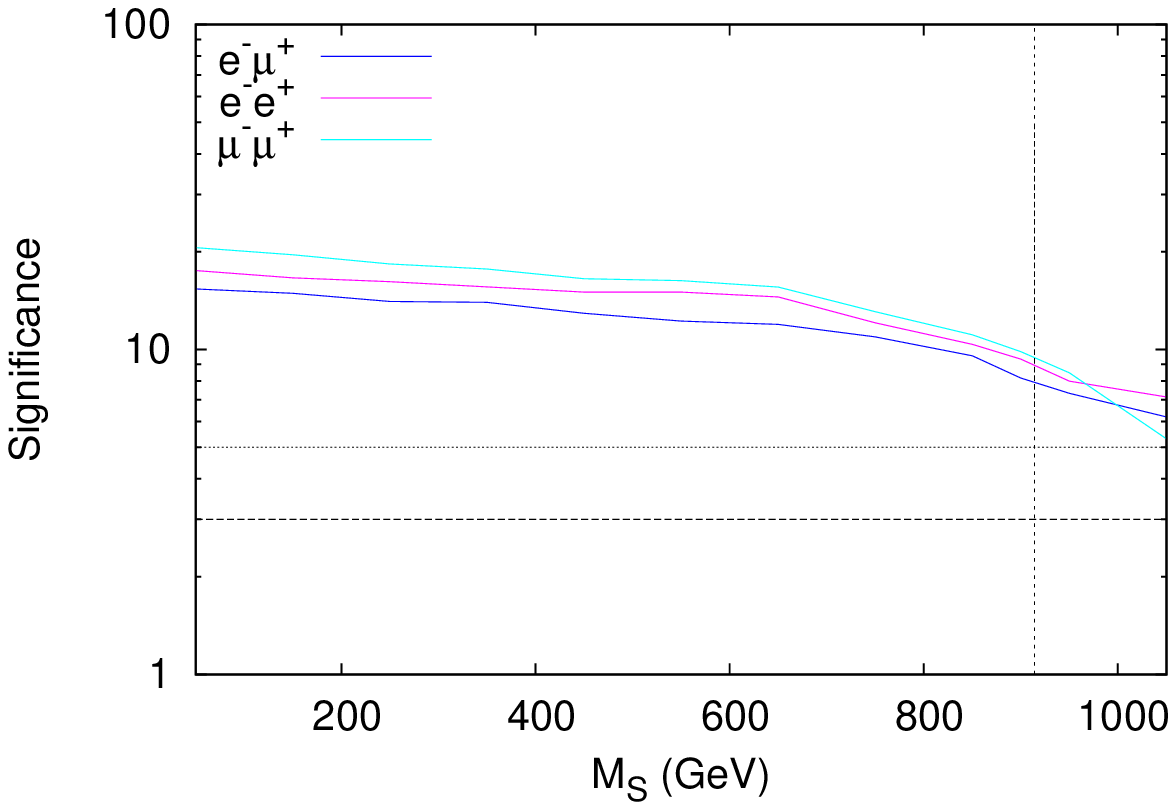}
\par\end{centering}
\caption{\textit{The significance versus the charged scalar mass, $M_{S_{1}}$,
at the CM energy 8 TeV (left) and 14 TeV (right). The two horizontal dashed
lines represent $\mathcal{S}=3$ and $\mathcal{S}=5$, respectively.}}%
\label{Slhc}%
\end{figure}

Turning now to the LHC, the processes $pp\rightarrow
e^{-}e^{+}+E_{miss}$, $\mu^{-}\mu^{+}+E_{miss}$,
$e^{-}\mu^{+}+E_{miss}$\ can be modified with respect to the SM,
where the missing energy could correspond to any of the combinations
mentioned above. We used CalcHEP \cite{calc} to generate different
distributions for two CM energies $E_{CM}=8,14$ TeV. After selecting
the cuts, we obtain the results in Fig.~\ref{Slhc},\ which shows the
significance versus the charged scalar mass $M_{S_{1}}$ for the
luminosity values $\mathcal{L}=20.3$ and 100 $fb^{-1}$ that
correspond $E_{CM}=8,14$ TeV, respectively~\cite{Guella}.

From Fig.~\ref{Slhc}-left, it is clear that the charged scalar mass should be
larger than 780 GeV, and from Fig.~\ref{Slhc}-right, we conclude that this
signal can be seen for LHC14.

\section{Conclusion}

We have shown that a generalized class of three-loop neutrino mass
models offers a promising way to experimentally probe the new
physics that is responsible for the origin of neutrino mass. We
showed that the models can solve both the neutrino mass and DM
problems without being in conflict with LFV constraints such as the
severe bound on $B(\mu\rightarrow e\gamma)$ and the muon anomalous
magnetic moment. We also investigated possible signatures at both
the LHC and ILC through the deviation from the SM in the processes
$pp\rightarrow e^{-}e^{+}+E_{miss}$, $\mu^{-}\mu^{+}+E_{miss}$,
$e^{-}\mu ^{+}+E_{miss}$ and $e^{-}e^{+}\rightarrow
e^{-}\mu^{+}+E_{miss}$ respectively. From the recent results of
LHC8, we put a bound on the charged scalar mass $M_{S_{1}}>780$ GeV.

\section*{Acknowledgments}

AA wants to thank the organizers for financial support. The authors thank R.
Soualah, C.~S.~Chen, T.~Toma, Ch.~Guella and D.~Chergui for productive
collaborations in this field.

\section*{References}


\end{document}